\documentclass{svjour3}                     % onecolumn (standard format)
\usepackage{graphicx}
\usepackage{amssymb}
\usepackage{float}
\usepackage{mathptmx}
\usepackage{amsmath}
\newcommand{\beq}{\begin{eqnarray}}
\newcommand{\eeq}{\end{eqnarray}}

\begin{document}

\title{Stealths on $(1+1)$-dimensional dilatonic gravity}

\titlerunning{Stealths on $(1+1)$-dimensional dilatonic gravity} % if too long for running head

\author{Abigail Alvarez            \and
        Cuauhtemoc Campuzano       \and
        Miguel Cruz                \and
        Efra\'\i n Rojas           \and
        Joel Saavedra
}


\institute{Abigail Alvarez \at
           Facultad de F\'\i sica, Universidad Veracruzana,
           91000, Xalapa, VER., M\'exico \\
           \email{abialvarez@uv.mx} \\
           \and
           Cuauhtemoc Campuzano \at
           Facultad de F\'\i sica, Universidad Veracruzana,
           91000, Xalapa, VER., M\'exico \\
           \email{ccampuzano@uv.mx} \\
           \and
           Miguel Cruz \at
           Facultad de F\'\i sica, Universidad Veracruzana,
           91000, Xalapa, VER., M\'exico \\
           \email{miguelcruz02@uv.mx} \\
           \and
           Efra\'\i n Rojas  \at
           Facultad de F\'\i sica, Universidad Veracruzana,
           91000, Xalapa, VER., M\'exico \\
           \email{efrojas@uv.mx} \\
           \and
           Joel Saavedra  \at
           Instituto de F\'\i sica, Pontificia Universidad Cat\'olica de Valpara\'iso,
           Casilla 4950, Valpara\'iso, Chile \\
           \email{joel.saavedra@ucv.cl}
           }


\maketitle

\begin{abstract}
We study gravitational stealth configurations emerging on a charged dilatonic
$(1+1)$-D black hole spacetime. We accomplish this by considering the coupling of
a non-minimally scalar field $\phi$ and a self-interacting scalar field $\Psi$
living in a  $(1+1)$-D charged black hole background. In addition, the
self-interacting potential for $\Psi$ is obtained which exhibits transitions for
some specific values of the non-minimal parameter. Atypically, we found that the
solutions for these stealth scalar fields do not have a dependence on the temporal
coordinate.

%\keywords{Stealth configuration \and Non-minimal coupling \and (1+1)-D black hole}
%\PACS{04.20.Jb \and 04.40.-b \and 04.50.Kd \and 04.60.Kz}

\end{abstract}

%%%%%%%%%%%%%%%%%%%%%%%%%%%%%%%%%%%%%%%%%%%%%%%%%%%%%%%%%%%%%%%%%%%%
\section{Introduction}
\label{sec:intro}
%%%%%%%%%%%%%%%%%%%%%%%%%%%%%%%%%%%%%%%%%%%%%%%%%%%%%%%%%%%%%%%%%%%

The Einstein equations are one of the pillars of the General Relativity (GR)
because they are quantifying the relationship between matter and the spacetime
curvature. The conventional GR prescription says that the curvature of the
spacetime is a manifestation of the presence of matter and this fact is encoded
in the Einstein equations
\begin{equation}
\emph{G}_{\mu\nu} + \Lambda g_{\mu\nu} - \kappa T_{\mu\nu} = 0,
\end{equation}
where $T_{\mu\nu}$ is the energy-momentum tensor associated to the matter content.
Needless to say that the finding physically interesting solutions to these
equations is a difficult task due to their high nonlinearity.

Nevertheless, there are nontrivial configurations without back reaction on the geometry
of the spacetime in which it has been shown the existence of a reverse reaction on the
geometry of spacetime. These fields are called \textit{stealths},~\cite{AyonBeato:2004ig}.
Such configurations are obtained from the Einstein equations when one impose the vanishing
of them including customary matter, and for the other hand, the vanishing of an
energy-momentum tensor associated to a non-minimally coupled scalar field, $\Psi$,
independently
\begin{equation}
0 = \emph{G}_{\mu\nu}+\Lambda g_{\mu\nu} - \kappa T_{\mu\nu} = \kappa T_{\mu\nu}^S=0.
\label{eq2}
\end{equation}
Originally it was shown the existence of these configurations in~\cite{AyonBeato:2004ig}
for the well known $(2+1)$-dimensional BTZ static black hole and later on in $(3+1)$-dimensions
for the Minkowski flat spacetime~\cite{AyonBeato:2005tu}. Thenceforth, this idea was
implemented in several spacetimes: in an (A)dS spacetime~\cite{AyonBeato:2005pr}, and in
the cosmology context~\cite{Banerjee:2006pr,Maeda:2012tu,Ayon-Beato:2013bsa}. Similarly,
in $n$ dimensions, for higher-order gravity theories~\cite{Gaete:2013ixa,Hassaine:2013cma}
in the Lovelock gravity context~\cite{Gaete:2013oda} as well as in the non-relativistic
version of the gauge/gravity correspondence in Lifshitz spacetime~\cite{Ayon-Beato:2015qfa}.
Surprisingly, a stealth Schwarzschild solution arise in context of Galileons theory,
in particular, on shift-symmetric part of Hordenski action, which is used to study
no-hair theorems in a spherically symmetric spacetime \cite{Babichev:2013cya} and its charged
generalization \cite{Babichev:2015haa}.
In all these works, the stealth scalar field arises from the well known non-minimal coupling
$\xi \Psi^2 R$ in general relativity. In contrast with the ordinary matter, it is believed that
the distinctive of these configurations is ability to be undetectable in the ambient spacetime
which makes it an interesting phenomenon which could explain some current phenomena of the
gravitation as in the case of dark matter/energy.
The mechanism described above is quite general and can be applied for different spacetimes.
Indeed, the current implementation of this idea on diverses geometries is under
investigation in order to get a good understanding of the physical content behind the stealth
setups. Despite the criticism on the instability of the stealth configurations, we believe
that this framework should not be ruled out. In fact, it has recently emerged a strong interest
in part because of the conformal symmetry properties and the potential cosmological implications
that have this type of structures~\cite{Ayon-Beato:2013bsa}.

In this work we will merely present, but without giving any compelling physical conclusion,
exact solutions for stealth configurations arising in the context of black hole physics for
lower dimensions. Specifically, such configurations are realized by considering
a non-minimally coupled scalar field, evolving in a two-dimensional charged dilatonic
black hole background~\cite{Frolov:1992xx}. Inspired in the string theory, this black
hole is very interesting in its own right due to their properties on the quantum effects
in the black hole physics and by the their implications in the quantization of
nonconformal fields, mainly. The solution for this geometry comes from string theory
in two dimensions where its structure is equivalent to the planar symmetry of general
relativity~\cite{Lemos:1994fn}. Due to the fact that the two-dimensional dilatonic black
hole satisfy Einstein equations, then it is also considered as a genuine black hole in
general relativity. When the exact solutions are obtained in this work, as a byproduct, one
is able to identify the self-interacting potential for the stealth scalar field. At the moment
there is no close relationship with any physical potential so it remains to be understood
the role that such potential play.

This paper is organized as follows. In Sect. 2 we review briefly the coupling between a
non-minimally scalar field with a $(1+1)$-D charged dilatonic black hole. In Sect. 3 we
obtain exact solutions for stealth configurations considering the two-dimensional charged
dilatonic black holes. In addition, we discuss some interesting cases for $\xi \neq 0$ where
stealth scalar field configurations appear. In all cases we were able to compute the accompanying
self-interacting potentials. Finally, in Sect.~\ref{conclu} we give some conclusions.

%%%%%%%%%%%%%%%%%%%%%%%%%%%%%%%%%%%%%%%%%%%%%%%%%%%%%%%%%%%%%%
\section{$(1+1)$-D charged dilatonic black hole coupled to a non-minimally a scalar field}
\label{sec:1}
%%%%%%%%%%%%%%%%%%%%%%%%%%%%%%%%%%%%%%%%%%%%%%%%%%%%%%%%%%%%%%

In this Section we review briefly the $(1+1)-$D gravity coupled
with a dilaton field $\phi$ and a electromagnetic field besides and additional term
involving a non-minimally coupled scalar field $\Psi$
\begin{eqnarray}
S%[g,\phi,\Phi]
&=& S_{\phi} + S_{\Psi},
\nonumber \\
&=& \frac{1}{2\kappa} \int d^{2}x\sqrt{-g} \,e^{-2\phi}
\left[ R+4\lambda^2 +4(\nabla_{\alpha}\phi\nabla^{\alpha}\phi)\right]
-\frac{1}{8\kappa} \int d^2 x \sqrt{-g} e^{-2\phi} F_{\mu\nu} F^{\mu \nu}
\nonumber
\\
&-&  \int d^{2}x\sqrt{-g} \left[ \frac{1}{2}\nabla_{\alpha}\Psi\nabla^{\alpha}\Psi
+\frac{\xi}{2}R\Psi^2 +U(\Psi)\right],
\label{action}
\end{eqnarray}
where $R$ denotes the scalar curvature, $\lambda$ is a constant, $\xi$ is the
non-minimal coupling parameter, $F_{\mu \nu}$ is the field strength of the Maxwell
field $A_\mu$ and $U(\Psi)$ represents the self-interacting potential for the scalar
field $\Psi$. Note that the action can be splitted in two parts, $S_{\phi}$ and $S_{\Psi}$,
that is, the gravitational and the scalar field sectors, respectively. The first
part corresponds to the $2D$ dilatonic gravity approach~\cite{Grumiller:2002}
which is the version of the Einstein-Hilbert theory in two dimensions with physical
degrees of freedom and the second part corresponds to the non-minimal coupling
of the scalar field $\Psi$. Furthermore, besides the constant $\kappa$, the quantity
$e^{\phi}$ plays the role of a gravitational coupling constant~\cite{Giddings:1992fp}.
Since we are interested in stealth configurations for the scalar field $\Psi$,
the quantity $e^{\phi}$ is not multiplying the second part of the action. In the
stealth scalar field framework, these non-trivial configurations do not produce backreaction.
These must be treated separately from the gravitational sector. The corresponding field
equations are obtained from the variation of the action (\ref{action}) with respect to the
associated fields
\begin{eqnarray}
&& \nabla_\mu\nabla_\nu \phi - g_{\mu\nu}\left(\Box \phi
-\nabla_\alpha \phi \nabla^\alpha \phi + \lambda^2
-\frac{1}{16}F_{\alpha\beta}F^{\alpha\beta}\right)
\nonumber
\\
&-& \frac{1}{4}F_\mu{}^{\alpha}F_{\alpha\nu}= \frac{\kappa}{2}
e^{2\phi}T_{\mu\nu}^{{}\textit{S}},
\label{eq:eq_Einstein21}
\\
&& R+4\Box \phi-4 \nabla_\alpha \phi \nabla^\alpha \phi
+4\lambda^2-\frac{1}{4}F_{\alpha\beta}F^{\alpha\beta} = 0,
\label{eq:mov2}
\\
&& \Box\Psi-\xi R \Psi = \frac{dU(\Psi)}{d\Psi},
\label{eq:mov3}
\\
&& \nabla_\mu(e^{-2\phi}F^{\mu\nu})=0.
\label{eq:mov4}
\end{eqnarray}
where the stealth's energy-momentum tensor reads
\begin{equation}
\label{eq:energy}
T_{\mu\nu}^{S} = \nabla_\mu \Psi \nabla_\nu \Psi - g_{\mu\nu}
\left( \frac{1}{2}\nabla_\alpha \Psi \nabla^\alpha \Psi + U(\Psi) \right) +
\xi( g_{\mu\nu} \Box - \nabla_\mu \nabla_\nu + G_\mu{}_\nu)\Psi^2.
\end{equation}
This expression is the key point in the search for stealth configurations.
We have to impose the vanishing condition of the l.h.s of the
Eq. (\ref{eq:eq_Einstein21}) as stated in Eq. (\ref{eq2}) in order to obtain the
stealth scalar field configurations.

%%%%%%%%%%%%%%%%%%%%%%%%%%%%%%%%%%%%%%%%%%%%%%%%%%%%%%%%%%%%%%%%%%%
\subsection{$(1+1)$-dimensional dilatonic dlack hole solution}
%%%%%%%%%%%%%%%%%%%%%%%%%%%%%%%%%%%%%%%%%%%%%%%%%%%%%%%%%%%%%%%%%%

In order to study the stealth scalar field configuration in lower dimensional
black hole geometries, we adopt the following background
\begin{equation}
\label{eq:metric}
ds^2 = -f_q(r)dt^2+\frac{dr^2}{f_q(r)}.
\end{equation}
This corresponds to a $(1+1)$-dimensional dilatonic black hole solution for the
Einstein equations described by the vanishing of l.h.s. of Eq. (\ref{eq:eq_Einstein21}).
When some specific coordinates are chosen, the function $f_q$ is explicitly given
by~\cite{Frolov:1992xx},
\begin{equation}
f_q (r) = 1 - \frac{2m}{Q} e^{-Q r} + \frac{q^2}{Q^{2}} e^{-2Qr},
\label{eq:fmetric}
\end{equation}
where $m$ is a mass parameter, $q$ is proportional to the charge, $Q$ is a positive
constant determined by the effective central charge $\lambda$ and the dilatonic field
can be written in the simple form
\begin{equation}
\phi = Qr.
\label{eq:phi-q}
\end{equation}
In this solution there is no contribution coming from a tachyon field and it was shown
in Refs.~\cite{McGuigan:1991qp,Witten:1991yr} that is the two dimensional version
of the four dimensional Schwarzschild black hole solution.

It is worthy to note that when $m\geq q$ the expressions~(\ref{eq:fmetric}) and~(\ref{eq:phi-q})
describes a charged $(1+1)$-dimensional black hole with charge $q$. When $q>m$, one gets
a naked singularity and for the specific choice $m=1/2, Q= 1/r_0$ and $q=0$, the function
$f_q(r)$ specializes to the well known uncharged case~\cite{Witten:1991yr,Sengupta,Frolov:2000jh},
\begin{equation}
\label{eq:funtion}
f(r) = 1 - r_{0}e^{-\frac{r}{r_0}}
\qquad
\phi=\frac{r}{r_0},
\end{equation}
where the parameter $r_0$ determines the position of the event horizon.

On the other hand, as mentioned previously, following~\cite{AyonBeato:2004ig,AyonBeato:2005tu}
the emerging of a stealth configuration is obtained by demanding the
vanishing of the r.h.s. of Eq.~(\ref{eq2}), bearing in mind that both sides of this
equation vanishes independently. Hence, we now turn to explore the conditions to obtain
non-trivial stealth scalar field configurations in this $(1+1)$-D charged black hole background.

%%%%%%%%%%%%%%%%%%%%%%%%%%%%%%%%%%%%%%%%%%%%%%%%%%%%%%%%%%%%%%%%%%%%%%%%%%%%%%%%
\section{Stealth configurations on a $(1+1)$-D charged dilatonic black hole}
\label{stealth}
%%%%%%%%%%%%%%%%%%%%%%%%%%%%%%%%%%%%%%%%%%%%%%%%%%%%%%%%%%%%%%%%%%%%%%%%%%%%%%

To begin with, we proceed to obtain a nontrivial solution for the scalar field
$\Psi$. It will be useful to rewrite this field as follows
\begin{eqnarray}
\label{eq:phi}
\Psi=\sigma^{\frac{2\xi}{4\xi - 1}},
\end{eqnarray}
where $\sigma=\sigma(t,r)$ is a function depending on the coordinates
and $\xi$ is the non-minimal coupling parameter satisfying $\xi \neq 1/4$.
Obviously, this ansatz tell us that the case $\xi = 1/4$ must be discussed
separately. Guided by the approach behind Eq.~(\ref{eq2}) all we need to do is to
demand the vanishing of $T_{\mu\nu}^{{}\textit{S}}=0$. By inserting~(\ref{eq:phi})
into (\ref{eq:energy}), the off-diagonal component and the difference between
the diagonal temporal and radial component of the energy-momentum tensor
associated to $\Psi$ reads
\begin{eqnarray}
\label{eq:Ttr}
T_r{}^t&=& -\frac{(2\xi)^2}{(4\xi-1)}\frac{\Psi^2}{\sigma}f_q ^{3/2}\,
\partial^2_{rt}(f_q ^{-1/2}\sigma),
\\
\nonumber\\
\label{eq:Trr_Ttt}
T_t{}^t-T_r{}^r &=&
\frac{4\xi^2}{(4\xi - 1)}\frac{\Psi^2}{\sigma}(f_q ^2\sigma_{rr}+\sigma_{tt}).
\end{eqnarray}
The fulfillment of these equations guarantee that the nonlinear
Klein-Gordon equation~(\ref{eq:mov3}) is automatically satisfied.
From the vanishing of Eq. (\ref{eq:Ttr}) we must note that the solution
for the sigma function can be recast as  a sum of independent functions as follows
\begin{equation}
\label{separable}
\sigma=\sqrt{f_q}T(t)+\alpha(r),
\end{equation}
accompanied with the condition
\begin{equation}
\label{dtt_drr}
\partial_t\left[\frac{\sigma (T_t{}^t-T_r{}^r)}{\Psi^2}\right]=
\frac{2\xi^2}{4\xi-1}\bigg[f_q \frac{d}{dr}\left(\frac{1}{\sqrt{f_q}}\frac{df_{q}}{dr}\right)
\frac{dT}{dt}
-2\frac{1}{\sqrt{f_q}}\frac{d^3 T}{dt^3}\bigg]=0.
\end{equation}
From Eqs.~(\ref{eq:Trr_Ttt}), (\ref{separable}) and (\ref{dtt_drr}) one is able to
show that the function $T$ is  a constant. This feature means that stealth scalar
field does not evolve in time. Thus, for $\xi\neq 1/4$ and considering $\sigma=\sigma(r)$
from the Eq.~(\ref{eq:Trr_Ttt}) one arrives at
\begin{equation}
\label{eq:s_field}
\Psi(r)= (\bar{A} r+ \bar{B})^{\frac{2\xi}{4\xi - 1}},
\end{equation}
where $\bar{A}$ and $\bar{B}$ are integration constants.
It is important to observe that this solution is determined independently of
the metric function and this scalar field coincides with the stealth
scalar field on the AdS background, as shown in~\cite{AyonBeato:2005pr}.

With this information we now attempt to find the self-interacting potential for
the scalar field $\Psi$. From the temporal-temporal component of the energy-momentum
tensor and by using the solution for scalar field~(\ref{eq:s_field}) and the
metric function~(\ref{eq:funtion}) we get the self-interaction potential for
$\xi \neq 1/4$ in the form
\begin{eqnarray}
U_\xi(\Psi)  &=&\bar{A}\frac{2\xi^2\Psi^{\frac{1}{2\xi}}}{Q^2 (1-4\xi)^2} \bigg\{
\left[2(1-4\xi)Q+  \bar{A} \Psi^{\frac{1-4\xi}{2\xi}}\right] q^2
\exp \left[ -2 Q\left(\frac{\Psi^{\frac{4\xi-1}{2\xi}} -
\bar{B}}{\bar{ A}}\right)\right]
\nonumber
\\
\nonumber
\\
&-&2 Q
\left[(1-4\xi)Q+  \bar{A} \Psi^{\frac{1-4\xi}{2\xi}}\right]  m
\exp\left[-Q\left(\frac{\Psi^{\frac{4\xi-1}{2\xi}}-\bar{B}}{\bar{ A}}\right)\right]
+ \bar{A}Q^2 \Psi^{\frac{1-4\xi}{2\xi}} \, \bigg \}.\quad
\label{eq:Ucharge}
\end{eqnarray}
This  potential allows the existence of stealth fields in $(1+1)$-D from the
metric~(\ref{eq:metric}). In Fig.~(\ref{fig:fig2}), for the sake of illustration,
we plot the expression (\ref{eq:Ucharge}) for the simple case $\xi = 1/2$ with
$\bar{A}$ just being a constant parameter. We can observe that the charged case
exhibits an absolute minimum near to the origin. This is due to the fact that the first
term dominates the behavior of the potential for small values
of $\Psi$.

%%%%%%%%%%%%%%%%%%%%%%%%%%%%%%%%%%%%%%%%%%%%%%%%%
\subsection{Some interesting cases}
%%%%%%%%%%%%%%%%%%%%%%%%%%%%%%%%%%%%%%%%%%%%%%%%%

%%%%%%%%%%%%%%%%%%%%%%%%%%%%%%%%%%%%%%%%%%%%%%%%%%%%%
\subsubsection{The $\xi=1/4$ case}
%%%%%%%%%%%%%%%%%%%%%%%%%%%%%%%%%%%%%%%%%%%%%%%%%%%%

\begin{figure}
\begin{center}
\includegraphics[width=60mm]{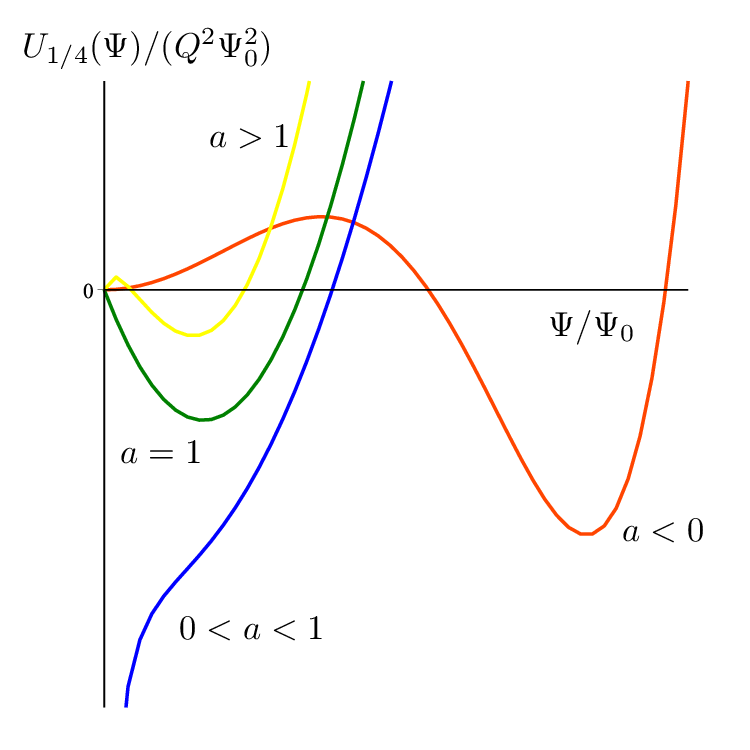}
\caption{This potential represents the self-interaction of the scalar field $\Psi$
for the value $\xi=1/4$. Here we have considered $\bar{\beta}=a Q$ and $q=m=Q/2 $,
where different values of $a$ has been taken, namely for the red line $a=-0.5$,
blue line $a=0.8$, the green $a=1$ and the yellow $a=1.5$.
\label{fig:fig1}}
\end{center}
\end{figure}

For this value, the solution~(\ref{eq:s_field}) is not longer valid therefore
we will focus on this specific case. By means of the redefinition of the scalar field
as follows
\begin{equation}
\Psi(t,r)= C e^{\sigma(t,r)},
\end{equation}
and inserting this together with the value $\xi=1/4$ in Eq.~(\ref{eq:energy}) we obtain
that the resulting equations can be reduced to
\begin{eqnarray}
\label{eq:Ttr_1_4}
T_r{}^t&=& -\frac{1}{2}\Psi^2f_q ^{3/2}\,
\partial^2_{rt}(f_q ^{-1/2}\sigma),
\\
\label{eq:Trr_Ttt_1_4}
T_t{}^t-T_r{}^r &=& \frac{1}{2}\Psi^2(f_q ^2\sigma_{rr}+\sigma_{tt}).
\end{eqnarray}
Following a similar procedure, it is straightforward to obtain the nontrivial solution
given by
\begin{equation}
\label{s14}
\Psi(r) = \bar{\Psi}_0 e^{\bar{\beta} r},
\end{equation}
where $\bar{\beta}$ and $\bar{\Psi}_0$ are integration constants. For this case the
self-interacting potential reads
\begin{eqnarray}
U_{1/4}(\Psi) &=&
\frac{m \Psi_0 ^2 (Q-2\bar{\beta})\bar{\beta}}{2Q}
\left(\frac{\Psi}{\Psi_0}\right)^{-\frac{(Q-2\bar{\beta})}{\bar{\beta}}}
-\frac{q^2 \Psi_0 ^2 (Q-\bar{\beta})\bar{\beta}}{2Q^2}
\left(\frac{\Psi}{\Psi_0}\right)^{-\frac{2(Q-\bar{\beta})}{\bar{\beta}}}
\nonumber \\
&+&\frac{\bar{\beta}^2}{2}\Psi^2.
\label{eq:Ucuarto}
\end{eqnarray}
The shape of the self-interaction potential~(\ref{eq:Ucuarto}) is depicted in
Fig.~(\ref{fig:fig1}). We infer that when $\Psi \rightarrow 0$, then $U(\Psi) \rightarrow 0$.
In fact, we have two minimums. On the other hand, the potential grows  as $\Psi$ is increased. In
addition, we must note the appearance of a local maximum.
%%%%%%%%%%%%%%%%%%%%%%%%%%%%%%%%%%%%%%%%%%%%%%%%%%%%%%%%%%%%%%%%%%%
%%%%%%%%%%%%%%%%%%%%%%%%%%%%%%%%%%%%%%%%%%%%%%%%%%%%%%%%%%%%%%%%%%%%

%%%%%%%%%%%%%%%%%%%%%%%%%%%%%%%%%%%%%%%%%%%%%%%%%%%%%%%%%%%%%%%
\subsubsection{The $\xi =1/2$ case}
%%%%%%%%%%%%%%%%%%%%%%%%%%%%%%%%%%%%%%%%%%%%%%%%%%%%%%%%%%%%%%%%

\begin{figure}
\begin{center}
\includegraphics[width=60mm]{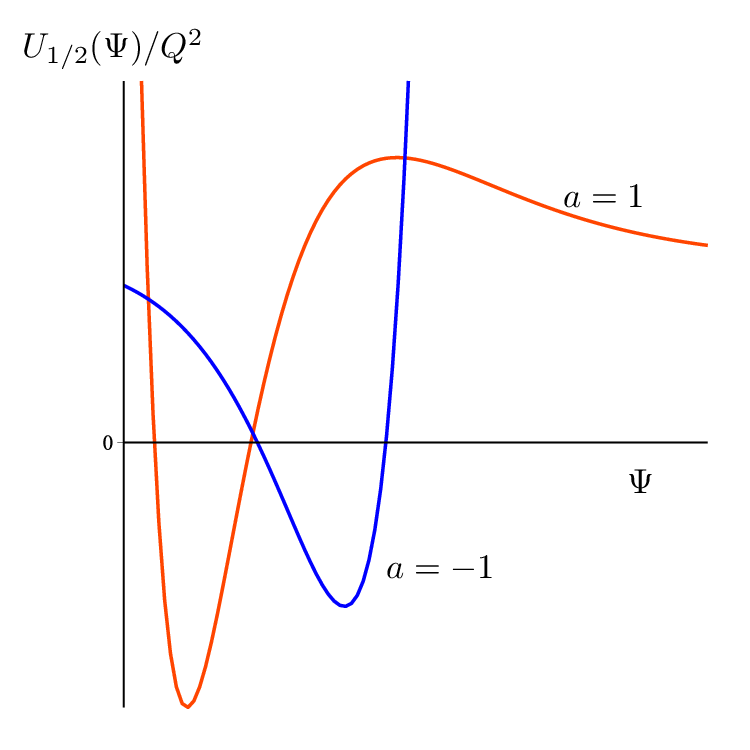}
\end{center}
\caption{ {Self-interaction potential for the simple case $\xi=1/2$ where
has been considered $a =\bar{A}/Q = -1$ (blue graph), $1$(red graph) and
$m/Q= 1/2$, $q/Q=1/2$ and $\bar{B}=2$. Remarkably, a local minimum appears
which is a generic fact for the charged case. See the expression (\ref{U:charge12})
below for more details.}
\label{fig:fig2}}
\end{figure}

For this case the associated solution is given by
\begin{equation}
  U_{1/2} = \frac{\bar{A}}{2Q^2} \left\{
\left( \bar{A} -2 Q \Psi \right) q^2 e^{\left[ -\frac{2Q}{\bar{A}}
\left( \Psi - \bar{B} \right)\right]} - 2Q
\left( \bar{A} - Q \Psi \right)  m \,e^{\left[-  \frac{Q}{\bar{A}} \left( \Psi - \bar{B} \right)\right]}
+\bar{A} Q^2 \right\}.
\label{U:charge12}
\end{equation}
%\begin{eqnarray}
%& U_{1/2} = \frac{\bar{A}}{2Q^2} \left\{
%\left( \bar{A} -2 Q \Psi \right) q^2 \exp \left[ -\frac{2Q}{\bar{A}}
%\left( \Psi - \bar{B} \right)\right] \right. \nonumber \\
%& \left. - 2Q
%\left( \bar{A} - Q \Psi \right)  m \exp \left[-  \frac{Q}{\bar{A}} \left( \Psi - \bar{B} \right)\right]
%+\bar{A} Q^2 \right\}
%\label{U:charge12}.
%\end{eqnarray}
From this expression we observe a linear dependence of the stealth field on the
self-interacting potential. In addition, the solution (\ref{eq:phi})
exhibits a linear dependence in the coordinate $r$ which leads to
a weird behaviour because the stealth fiels will permeate the space as $r$
grows. This potential is plotted in~(\ref{fig:fig2}).
%%%%%%%%%%%%%%%%%%%%%%%%%%%%%%%%%%%%%%%%%%%%%%%%%%%%%%%%%%%%%%%%%%%%%%%
%%%%%%%%%%%%%%%%%%%%%%%%%%%%%%%%%%%%%%%%%%%%%%%%%%%%%%%%%%%%%%%%%%%%%%

%%%%%%%%%%%%%%%%%%%%%%%%%%%%%%%%%%%%%%%%%%%%%%%%%%%%%%%%%%%%%%%%%%%
\subsubsection{The uncharged case}
%%%%%%%%%%%%%%%%%%%%%%%%%%%%%%%%%%%%%%%%%%%%%%%%%%%%%%%%%%%%%%%%%%%

\begin{figure}
\begin{center}
\includegraphics[width=60mm]{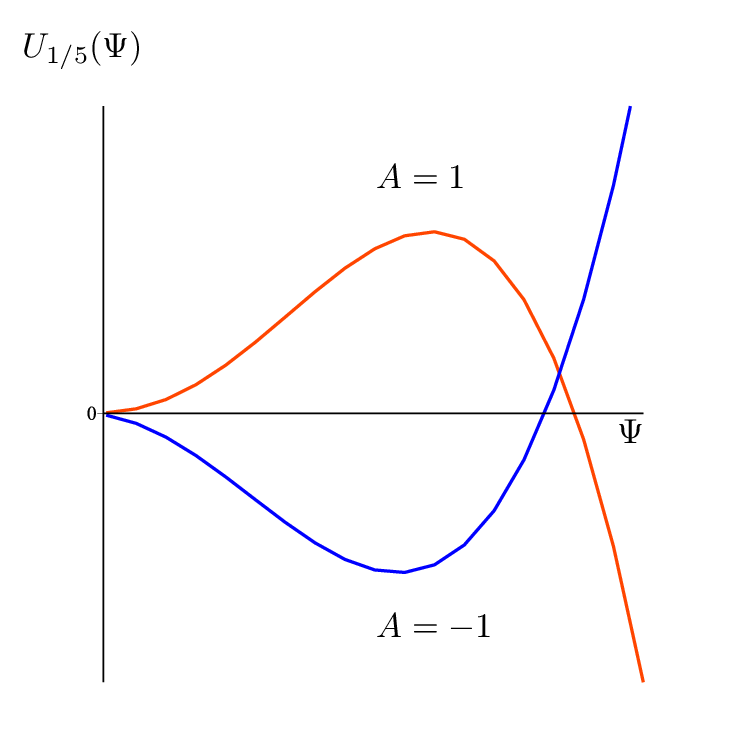}
\caption{These graphs represent the self-interaction potential.
In the figure, the red graph represent the self-interaction potential
with  constants set to the values $A=1$.
The blue graph represent the self-interaction potential
for the values $A=-1$. In both cases $r_0=1$ and $B =1/2$.
\label{fig:fig3}}
\end{center}
\end{figure}

When the electromagnetic field is switched off the nontrivial solutions found
previously specialize to the uncharged case straightforwardly. In this case
the equations of motion are modified slightly if we consider the vanishing
of $F_{\mu \nu}$ besides $q=0$ in the  geometric solution given by~(\ref{eq:funtion}).
Explicitly, the set of solutions are provided by the self-interacting potential
\begin{eqnarray}
U_\xi(\Psi) = {A}\frac{2\xi^2}{(1-4\xi)^2} \Big\{ &-&
\left[  (1-4\xi) +A r_0 \Psi^{\frac{1-4\xi}{2\xi}}\right]
\,e^{ - \frac{\left( \Psi^{\frac{4\xi-1)}{2\xi}} -
B\right)}{A r_0}}
+
A \Psi^{\frac{1-4\xi}{2\xi}} \Big \}\Psi^{\frac{1}{2\xi}}.\qquad
\label{U:uncharge}
\end{eqnarray}
Here, we have considered that $m=1/2$, $Q=1/r_0$ where $A, B$ are constant parameters.
For the sake of illustration, the self-interacting potential for the particular value
$\xi = 1/5$ is plotted in Fig.~(\ref{fig:fig3}). We can observe that in this case the
this potential exhibits a local minimum or a local maximum for different values of the
parameter $a:=A r_0$.
%%%%%%%%%%%%%%%%%%%%%%%%%%%%%%%%%%%%%%%%%%%%%%%%%%%%%%%%%%%%%%%%%%%%%%%%
%%%%%%%%%%%%%%%%%%%%%%%%%%%%%%%%%%%%%%%%%%%%%%%%%%%%%%%%%%%%%%%%%%%%%

As in the charged configuration situation, there is a special value for the
non-minimal coupling parameter, $\xi=1/4$, namely. It is necessary a separate
treatment in order to obtain a nontrivial solution. Since the mechanism already
performed is quite general in such case the stealth solution reduces to
\begin{equation}
 \Psi (r) = \Psi_0 \,e^{\beta r},
\end{equation}
where $\alpha$ and $\Psi_0$ are integration constants while its associated
self-interacting potential reads
\begin{equation}
U_{1/4} = \frac{\beta}{4} \left[ (1-2\beta r_0) \left(
\frac{\Psi}{\Psi_0} \right)^{-\frac{1}{\beta r_0} } + 2\beta \right] \Psi^2.
\end{equation}

%%%%%%%%%%%%%%%%%%%%%%%%%%%%%%%%%%%%%%%%%%%%%%%%%%%%%%%%%%
\section{Conclusions}
\label{conclu}
%%%%%%%%%%%%%%%%%%%%%%%%%%%%%%%%%%%%%%%%%%%%%%%%%%%%%

In this work we have obtained the stealth scalar field configurations arising
in a $(1+1)$-dimensional dilatonic black hole spacetime. We found
that the solutions for the stealth scalar field do not evolve in time.
These solutions were obtained using the most general metric
ansatz~(\ref{eq:metric}). It is worth noting that these solutions
look similar to those obtained for the case of the stealth field on the
AdS background, as shown in~\cite{AyonBeato:2005pr}. The simplicity of the mathematical
manipulations in $(1+1)$-dimensional spacetimes and the similarities that they possess with
the four dimensional case, provide some hints to glimpse the properties of more complicated
cases, in this sense, we expect a similar behavior for the stealth configurations that
could appear if the Schwarzschild solution is used as background. Needless to say, a number
of theoretical questions are still to be addressed.
Most notably, it is no clear what degrees of freedom are represented by this
scalar field. Work is in progress.
For example, with regards the potential function, a more detailed analysis
on the integration constants and the values of $\xi$ is mandatory in order
to have an acceptable behaviour. In addition, the relationship of this
potential with a known physical system does not exist yet.

On the other hand, the behavior of the stealth will depend strongly on the value of
the exponent of $\Psi$ (see  Eqs.~(\ref{eq:s_field}) and~(\ref{s14})). For example,
for the exponent  $\Xi:=(4\xi-1)/(2\xi)$ in the range $1< \Xi <\infty$
the stealth has positive exponents of $r$, and the behaviour seems like
a typical polynomial function. As mentioned in the text, the particular value
$\xi = 1/2$ casts out  a linear dependence of the stealth in the coordinate $r$
which leads to a unbounded behaviour because the stealth will permeate the space as
$r$ grows .
To end this discussion it is necessary to be cautious with the rational values of $\Xi$
in order to have a real value for $\Psi$. This can be achieved with an acceptable
choice of the integration constants $\bar{A}$ and $\bar{B}$.
Finally, we expect the existence of stealth solutions
considering another type of non-minimal couplings, like $\xi G^{\mu\nu}
\nabla_\mu \Psi \nabla_\nu \Psi$
\cite{Sushkov:2009}, by applying this procedure. It will be reported elsewhere.

%%%%%%%%%%%%%%%%%%%%%%%%%%%%%%%%%%%%%%%%%%%%%%%%%%%%%%%%%%%%%%%%%
\begin{acknowledgements}
The authors acknowledge to Eloy Ay\'on-Beato for enlightening discussions. AA
acknowledges pasrtial support by CONACyT Grant Estancias Posdoctorales
Vinculadas al Fortalecimiento de Calidad del Posgrado Nacional 2016-1.
AA, CC and ER acknowledges partial support by CONACyT Grant CB-2012-177519-F and
grant PROMEP, CA-UV, \'Algebra, Geometr\'{\i}a and Gravitaci\'on. CC also
acknowledges CONACyT Grant I0010-2014-02 Estancias Internacionales-233618-C.
MC is being supported by CONACYT- M\'exico through grant Repatriaciones 2015-04.
JS thank the hospitality of Facultad de F\'{\i}sica Universidad Veracruzana.
This work was partially supported by SNI (M\'exico).
\end{acknowledgements}
%%%%%%%%%%%%%%%%%%%%%%%%%%%%%%%%%%%%%%%%%%%%%%%%%%%%%%%%%%%%%%%%%%%%%%

%\appendix


\end{document}